\documentclass[%
 reprint,
 amsmath,amssymb,
 aps,
]{revtex4-2}

\usepackage{graphicx}% Include figure files
\usepackage{dcolumn}% Align table columns on decimal point
\usepackage{bm}% bold math
\usepackage{amsmath} %per scrivere matematica
\usepackage{amssymb} %per simboli matematici
\usepackage{amsthm} %per format. i thm/def/lemma
\usepackage{bbm} %per lettere/num (tipo N,1,...R)
\usepackage[usenames,dvipsnames]{xcolor} %per i colorare il testo
\usepackage{cancel} %per fare le semplificazioni
\usepackage[shortlabels]{enumitem} %per formattare enumerate
\usepackage{float}
\usepackage{verbatim}
\usepackage[caption=false]{subfig}
\newcommand{\E}{\mathbb{E}}
\renewcommand{\P}{\mathbb{P}}

\newcommand{\bx}{x}

\newcommand*{\swap}[2]{#2#1}

\newcommand{\Exp}[1]{\mathbb{E}\left[#1\right]}
\newcommand{\Prob}[1]{\mathbb{P}\left(#1\right)}
\newcommand{\Var}{\operatorname{Var}}
\newcommand{\Cov}{\operatorname{Cov}}

\begin{document}

%\preprint{APS/123-QED}

\title{Detecting hyperbolic geometry in networks: \\
why triangles are not enough}% Force line breaks with \\

\author{Riccardo Michielan}
\author{Nelly Litvak}
\author{Clara Stegehuis}
\affiliation{%
 Faculty of Electrical Engineering, Mathematics and Computer Science, University of Twente }%

\date{November 8, 2022}% It is always \today, today,

\begin{abstract}
In the past decade, geometric network models have received vast attention in the literature. These models formalize the natural idea that similar vertices are likely to connect. Because of that, these models are able to adequately capture many common structural properties of real-world networks, such as scale invariance and high clustering. Indeed, many real-world networks can be accurately modeled by positioning vertices of a network graph in hyperbolic spaces. Nevertheless, if one observes only the network connections, the presence of geometry is not always evident. Currently, triangle counts and clustering coefficients are the standard statistics to signal the presence of geometry. In this paper we show that triangle counts or clustering coefficients are insufficient because they fail to detect geometry induced by hyperbolic spaces. { We therefore introduce a novel statistic, \textit{weighted triangles}, which weighs triangles based on their evidence for geometry. We show analytically, as well as on synthetic and real-world data, that weighted triangles are a powerful statistic to detect hyperbolic geometry in networks.}
\end{abstract}
\maketitle

%\tableofcontents

\section{Introduction}
Network geometry is one of the most important features of real-world networks~\cite{boguna2021}, as it explains frequently observed network properties, including scale-invariance, high clustering, and overlapping community structures. The concept of geometry is intimately connected to the idea of similarity in networks, that two similar vertices are likely to connect. Typical examples are social networks, where people sharing the same interests or living in the same region are more likely to interact, or the world wide web, where webpages sharing similar contents are often linked. The similarity of vertices can be modeled through geometry in a random graph. More specifically, two individuals can be modeled as vertices in a geometric space, and their similarity is then associated with a small distance between them. This class of models is often called  geometric network models.

In the past decades, several geometric models have been introduced, aiming to model and understand real network properties~\cite{krioukov2010,bringmann2016,deijfen2013scale,serrano2008}. The simplest example of geometric models are the random geometric graphs \cite{penrose2003,dall2002}, where an underlying Euclidean space is fixed, and nodes within a given radius are connected. However, random geometric graphs fail to describe the typical scale-free nature of real-world networks, where many nodes of low degree coexist with a few nodes of extremely large degrees. %Therefore in most cases, euclidean spaces yield inaccurate models to study complex networks. 
Instead, the hyperbolic space has been found to be extremely useful for modeling real-world networks~\cite{krioukov2010}, as it explains the heavy-tailed degree distributions that many networks posses, their efficient routing~\cite{blasius2018,kleinberg2007} and their community structures~\cite{faqeeh2018}. Interestingly, the hyperbolic model is equivalent to the renowned $\mathbb{S}^1$ model, where nodes are uniformly distributed on the circle, and a soft constraint is assigned to the degrees~\cite{serrano2008}. 
Recently, a very intuitive network model called geometric inhomogeneous random graph (GIRG) has been proposed as a generalization of hyperbolic random graphs~\cite{bringmann2016}. GIRGs combine the essential real-world properties of the hyperbolic random graph with simpler mathematical formulations that make them mathematically more tractable than hyperbolic random graphs. In these models each vertex is positioned in a geometric space, and has a weight representing its ability to attract connections. Then, the probability of a connection between two vertices in the GIRG is an increasing function of the weight of each vertex, and a decreasing function of a distance between them.  

Although in many cases network geometry is a natural assumption, networks may not be directly endowed with a metric space, and in principle they could also not posses any kind of geometry.
Determining whether a given network has an underlying latent hyperbolic geometric space has been highlighted as a major open problem~\cite{boguna2021,smith2019}. Indeed, usually we can only observe the network connections, and not the network features which can cause geometry. The main question is then: can we assess whether these connections were formed by some underlying geometry? 

It has often been assumed that high clustering, or a large amount of triangles, implies the presence of geometry~\cite{krioukov2016}. Intuitively,  in a geometric space, the triangle inequality ensures that two neighbors of a vertex are likely to be close to each other, and therefore likely to be connected. In fact, triangles are a powerful statistic to distinguish geometric and non-geometric networks, when assuming that the former can be endowed with an Euclidean space~\cite{devroye2011,gao2017,bubeck2016}. %On the other hand, hyperbolic models show high average clustering \cite{fountoulakis2021,GIRG}.
In this paper however, we show that pure triangle counts, as well as the average clustering coefficient, have limitations in detecting geometry induced by hyperbolic spaces. In fact, network models without any source of geometry may contain the same amount of triangles as hyperbolic random graphs, and may have arbitrary large average clustering coefficients. This is essentially due to the trade-off between weights and geometry in the GIRG representation of the hyperbolic model: connections can either be formed between high-weight nodes, or between close--by nodes. When the degree distribution has a particularly heavy tail, this trade-off favors triangles formed by high weighted vertices. Therefore, the underlying geometry cannot always be detected by triangle counts or clustering coefficients.

In this work we develop a novel and powerful statistic called \emph{weighted triangles} to detect hyperbolic geometry in networks, where triangles are weighted based on how likely they are to be caused by geometry. Triangles that carry low evidence for geometry are discounted, so that our statistic is able to differentiate hyperbolic geometry from non-geometric networks in a regime where standard triangle counts or clustering coefficients are not~\cite{kolk2021,michielan2021}. 
We also provide experimental evidence that weighted triangles can successfully detect a geometric nature of real-world networks.%, in contrast to both triangle counts and average clustering coefficient.
\\

{
\paragraph*{Notation.}
%In this paper we consider large networks, where the number of vertices $n$ is very large. In particular, the presented results on pure triangle counts, average clustering coefficient, and weighted triangles hold asymptotically in terms of $n$. Then, it is convenient to introduce notations to express asymptotic relations, as $n$ goes to infinity.
We say that sequence of events $\{\mathcal{E}_n\}$ occurs with high probability (w.h.p.) when
%the probability of $\mathcal{E}_n$ goes to 1 as $n$ goes to infinity, that is, 
$\lim_{n \to \infty} \P(\mathcal{E}_n) = 1$. 
We say that a sequence $\{X_n\}$ of random variables converges to the random variable $X$ in probability, $X_n \stackrel{\mathbb{P}}{\longrightarrow} X$, when $\lim_{n\to \infty } \mathbb{P} (|X_{n}-X|>\varepsilon) = 0$.
}

\section{Models}
%Hidden-variable models form a class of popular null models for scale-free networks, due to their simplicity and their ability to capture real-world network properties such as heavy-tailed degree distributions. In their most simple form, hidden-variable models do not contain geometry. However, they can be enriched by adding latent variables that describe the positions of the nodes. This makes hidden-variable models a natural choice to distinguish geometric and non-geometric networks and obtain a precise characterization of suitable statistics to detect hyperbolic geometry.

In this paper, our benchmark model is the Geometric Inhomogeneous random graph~\cite{GIRG}, denoted by GIRG. This is a powerful, but tractable, random graph model that generalizes the hyperbolic random graph defined by Krioukov et al. in~\cite{krioukov2010}. In the GIRG model, each vertex $i\in \{1,2,\ldots,n\}$ is equipped with a weight $h_i$ and a uniformly sampled position $x_i$ on a $d$-dimensional torus $[0,1]^d$ endowed with the infinity norm. Weights are sampled independently from the Pareto distribution, with density
\begin{equation}\label{eq:paretolaw}
    \rho(h)=K_1 h^{-\tau},
\end{equation}
for any $h > h_0 > 0$, where $K_1$ is the normalizing constant and $\tau\in(2,3)$. Two nodes $i$ and $j$, are then connected independently with probability 
\begin{equation}\label{eq:edgeprobGIRG}
    p(h_i,h_j,x_i, x_j)=K_2 \min\left(\frac{h_ih_j}{\mu n} ||x_i - x_j||^{-d},1\right)^{\gamma},
\end{equation}
for some $\gamma>1$ and where $\mu$ denotes the average weight. The constant $K_2$ is a correction factor, which ensures that the expected degree of any vertex $i$ is proportional to its weight $h_i$ (see Appendix \ref{app:expgedreeGIRG}).

In its original formulation, Bringmann et al. \cite{bringmann2016} defined the GIRG as an entire class of models, where the connection probability \eqref{eq:edgeprobGIRG} is only required to hold asymptotically. Here we focus on a specific element of this class, to have a practical and tractable comparison with non-geometric models. 
An important reason to study the GIRG is that for $d=1,\gamma=\infty$ it is asymptotically equivalent to the hyperbolic random graph, as shown in \cite{GIRG,komjathy2020}. Thus, the asymptotic results on the GIRG shown here are valid for the hyperbolic random graph as well. 

To determine when geometry can be detected, we compare the GIRG to the Inhomogeneous random graph, which we denote by IRG. This model differs from the GIRG only in the sense that it is non-geometric. Specifically, the vertices now possess only weights $h_i$, sampled from the Pareto distribution as in \eqref{eq:paretolaw}, and are connected with probability
\begin{equation}\label{eq:edgeprobIRG}
    p(h_i,h_j)=\min\left(\frac{h_i h_j}{\mu n},1\right).
\end{equation}
The two models have the same weight, and degree distribution \cite{bollobas2007,hofstad2021b}, and  differ only in their connection probability through the presence of geometry. We will therefore show that under classical triangle-based statistics, these models cannot be distinguished, making it impossible to distinguish hyperbolic geometry from simple scale-free networks.

\section{Triangle counts}
%It has often been assumed that high clustering, or a large amount of triangles, implies the presence of geometry~\cite{krioukov2016}. Indeed, many existing statistics to infer the presence of geometry use triangles~\cite{devroye2011,gao2017,bubeck2016}. 
%However, we show now that standard measures of clustering, such as triangle counts or clustering coefficients, cannot be used to detect hyperbolic geometry. 

\begin{figure*}[htb]
\centering
\subfloat[$\tau = 2.9$ \label{fig:pure_triangles_models_large}]{\includegraphics[width=0.4\textwidth]{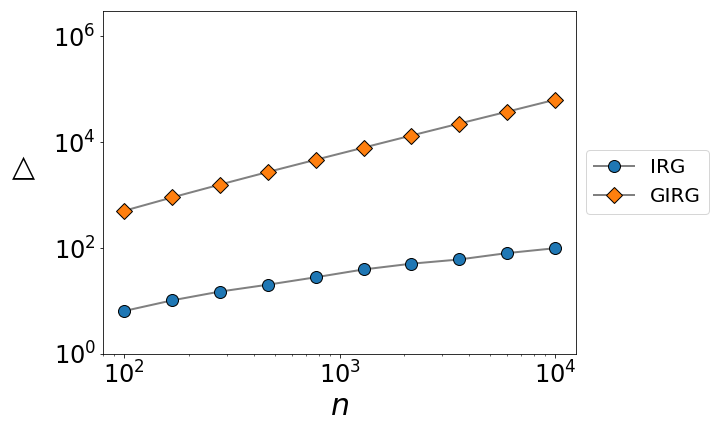}}
\qquad
\subfloat[$\tau = 2.1$ \label{fig:pure_triangles_models_small}]{\includegraphics[width=0.4\textwidth]{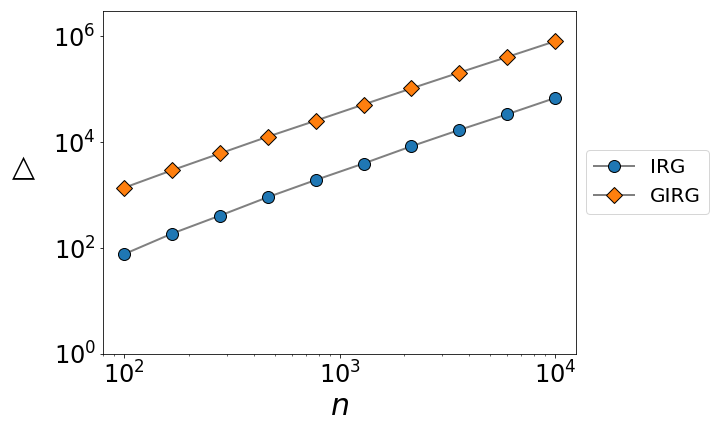}}

\subfloat[$\tau = 2.9$ \label{fig:average_clustering_models_large}]{\includegraphics[width=0.4\textwidth]{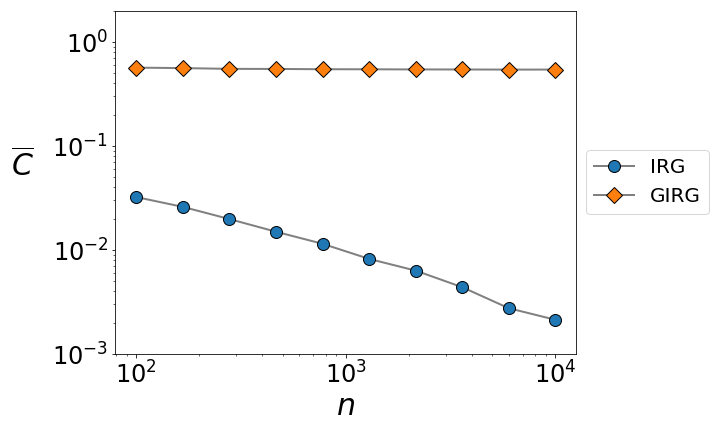}}
\qquad
\subfloat[$\tau = 2.1$ \label{fig:average_clustering_models_small}]{\includegraphics[width=0.4\textwidth]{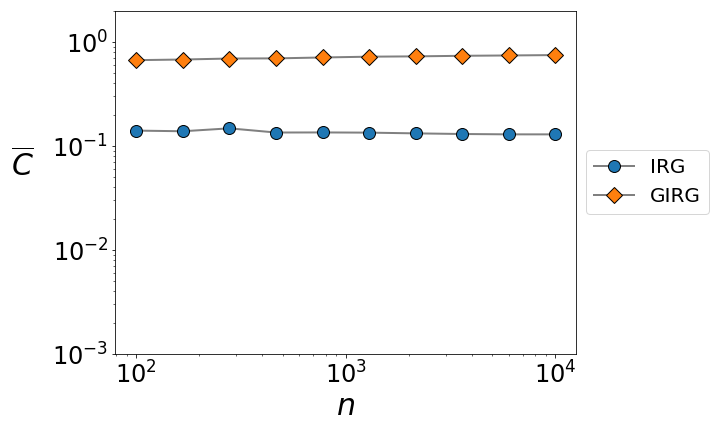}}

\subfloat[$\tau = 2.9$]{\includegraphics[width=0.4\textwidth \label{fig:weighted_triangles_models_large}]{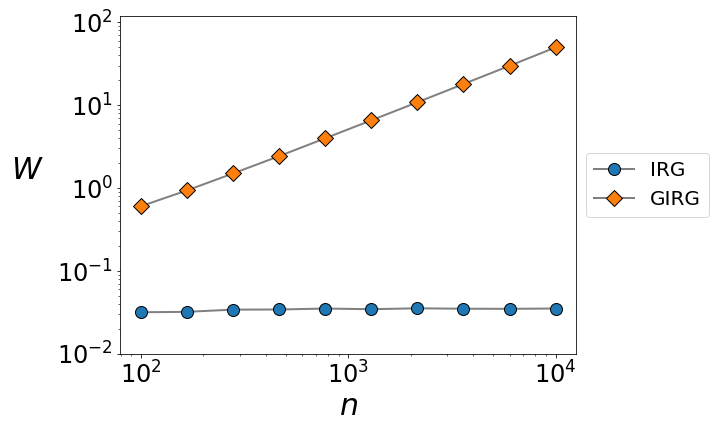}}
\qquad
\subfloat[$\tau = 2.1$ \label{fig:weighted_triangles_models_small}]{\includegraphics[width=0.4\textwidth]{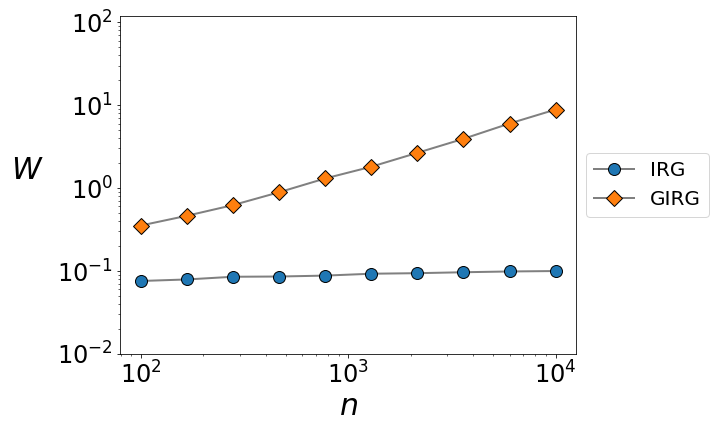}}

\caption{Triangle counts, clustering coefficient and weighted triangles for the Inhomogeneous random graph (IRG) and Geometric Inhomogeneous random graph (GIRG). Dots are the averages over 100 samples of IRG and GIRG, against the number of vertices $n$. Grey lines are linear interpolations between dots. (a) When $\tau>7/3$, the number of triangles increases linearly in GIRG, but slower than linearly for IRG. (b) When $\tau <7/3$, the triangles in the non-geometric and geometric model share the same scaling. (c) When $\tau$ is large, $\overline{C}$ differs significantly between the non-geometric and geometric model. (d)  When $\tau$ is small, $\overline{C}$ is qualitatively similar in both models. (e) (f) $W$ is significantly different both for small and large values of $\tau$. In particular, in non-geometric models $W$ remains bounded below 1/6.}

\end{figure*}

The most intuitive statistic to detect geometry is the triangle count in the network, $\triangle$. Indeed, one would expect a high number of triangles in geometric networks due to the triangle inequality.
The number of triangles in the IRG scales as $n^{3(3-\tau)/2}$~\cite{stegehuis2019}. In the GIRG on the other hand, the triangle count undergoes a phase transition: when $\tau < 7/3$ it scales as $n^{3(3-\tau)/2}$, and when $\tau > 7/3$ it scales as $n$~\cite{hofstad2021,blasius2018b,michielan2021}. Then, for small values of $\tau$, we cannot distinguish geometry from the triangle counts scaling, as in both models the number of triangles is proportional to $n^{3(3-\tau)/2}$.
In Figures \ref{fig:pure_triangles_models_large} and~\ref{fig:pure_triangles_models_small} we clearly see that when $\tau = 2.1$, the number of triangles in geometric and non-geometric models grow with the same scaling.

Theoretically, even when $\tau$ is small, it could still be possible to distinguish geometric/non-geometric inhomogeneous models through triangle counts. Indeed, if the number of triangles is $\triangle = A n^{3(3-\tau)/2} (1 +o(1))$, we may be able to establish the correct underlying model, by identifying the leading order term $A$. To this end, we can use precise asymptotic results from our earlier work~\cite{stegehuis2019,michielan2021}:

In the IRG, \begin{equation*}
    \frac{\triangle}{n^{3(3-\tau)/2}} \stackrel{\P}{\longrightarrow} A_{\text{IRG}}.
\end{equation*}

In the GIRG, when $\tau < 7/3$, \begin{equation*}
    \frac{\triangle}{n^{3(3-\tau)/2}} \stackrel{\P}{\longrightarrow} A_{\text{GIRG}}.
\end{equation*}

Here, $A_{\text{IRG}}$ and $A_{\text{GIRG}}$ are explicit constants that  depend on the parameters of the two models. 
If these are all known, then the two models could be distinguished using the difference between $A_{\text{IRG}}$ and $A_{\text{GIRG}}$. However, in realistic situations, one observes a network only through its connections, while the possible underlying geometry of the network and the vertex weights are usually not known. Therefore, in practice, we cannot estimate the parameters of the underlying models, and thus it is impossible to establish whether the triangle count is close to $A_{\text{IRG}} n^{3(3-\tau)/2}$ or $A_{\text{GIRG}} n^{3(3-\tau)/2}$. 

We conclude that the number of triangles is not a satisfactory statistic for inferring the hyperbolic geometry.

\section{Average clustering coefficient}
Another common network statistic to measure the tendency of the nodes to form triangles, is the clustering coefficient, which has also been ascribed to indicate a network geometry~\cite{krioukov2016}. 

The average clustering coefficient captures the probability that two randomly chosen neighbors of a randomly chosen vertex are connected as well. Equivalently, it is the average fraction of realized triangles out of all possible triangles that include a randomly chosen vertex: 
\begin{equation}
    \overline{C} = \frac{1}{n} \sum_{i \in V} C_i = \frac{1}{n} \sum_{i \in V} \frac{2 \triangle_i}{d_i(d_i-1)},
\end{equation}
where $\triangle_i$ is the number of triangles containing vertex $i$, and $d_i$ is the degree of vertex $i$. Typically, if the value of $\overline{C}$ does not vanish in $n$, then this is considered to be evidence for geometry. 

The average clustering in GIRGs is non-vanishing as $n$ increases~\cite{GIRG}. In particular, in hyperbolic random graphs, $\overline{C}$ converges in probability to a positive constant~\cite{fountoulakis2021}.
For the IRG on the other hand, $\overline{C}$ decays asymptotically, as $n^{\tau-2}\ln(n)$~\cite{colomer2012,hofstad2017b}. Even though $\overline{C}$ vanishes for large $n$, the decay of $\overline{C}$ is extremely slow when $\tau \approx 2$, as we can observe in Figure \ref{fig:average_clustering_models_large}. Thus, the average clustering coefficient is impractical to detect geometry when $\tau$ is small.

% \begin{figure*}[ht]
% \centering

% \caption{Average clustering coefficient in IRG and GIRG. (a) When $\tau$ is large, the behaviour of $\overline{C}$ differs significantly between the non-geometric and geometric model. (b)  When $\tau$ is small, $\overline{C}$ is qualitatively similar in both models.}
% \label{fig:average_clustering_models}
% \end{figure*}

Even more importantly, in non-geometric models with a different heavy-tailed degree distribution, $\overline{C}$ can even be constant. For example, consider an inhomogeneous random graph with weights
\begin{equation*}
    h_i=\begin{cases}
    2, & \text{with probability }1-1/(\sqrt{\mu n}),\\
    \sqrt{\mu n}, & \text{with probability } 1/(\sqrt{\mu n}).
    \end{cases}
\end{equation*}
We show in Appendix \ref{app:2-type_model} that $\overline{C}$ is constant in this IRG, even though the model does not contain any source of geometry. Hence, $\overline{C}$ too, is not a good statistic to distinguish between geometric and non-geometric networks.

\section{Weighted triangles}
We will now describe our proposed statistic that, contrary to standard clustering-based statistics, is able to detect hyperbolic geometry. This statistic is again triangle-based so that it has the same computational complexity as other clustering-based statistics. The difference between  our statistic and  the average clustering coefficient or triangle count, is that all triangles are weighted based on their evidence for geometry. Indeed, in a hyperbolic space, triangles can be formed because of popularity (high-degree vertices), or similarity (close--by vertices). Thus, we weigh each triangle, so that triangles which carry low evidence for geometry have low weight, and triangles which carry high evidence for being formed due to geometry have high weight. 
More precisely, we define the \textit{weighted triangles} count
\begin{equation}
    W := \sum_{\substack{i,j,k \in V \\ i<j<k}} \frac{1}{d_i d_j d_k}\mathbbm{1}_{\{(i,j,k) = \triangle\}},
\end{equation}
where $\mathbbm{1}_{\{(i,j,k) = \triangle\}}$ is the indicator function of the event that the vertices $i,j,k$ form a triangle in the network.
The intuition behind the weights $(d_id_jd_k)^{-1}$ is that, in non-geometric power-law models, a typical triangle is formed between vertices of high degrees \cite{hofstad2021}. Thus, when we \textit{discount} the triangles formed by high-degree vertices, $W$ remains small. On the other hand, in the GIRG, there is a large number of triangles formed by close--by vertices with small degrees. Therefore, $W$ is significantly greater for GIRGs than for IRGs.

Indeed, as we show in Appendix \ref{app:concentrationW}, w.h.p.,
\begin{align*}
    W &\leq 1/6, \text{ in IRG},\\
    W &\geq \delta n, \text{ in GIRG},
\end{align*}
for some $\delta > 0$.
Thus, weighted triangles make a very powerful statistic to infer hyperbolic geometry: in the absence of geometry, the statistic remains bounded, while it scales at least linearly under a geometric model. Moreover, this different behaviour is independent of $\tau$. This is clearly seen in Figures \ref{fig:weighted_triangles_models_large} and~\ref{fig:weighted_triangles_models_small}, where $W$ shows different scaling in non-geometric/geometric models, regardless of $\tau$. 

% \begin{figure*}[ht]
% \centering

% \caption{Weighted triangles in the two models. $W$ is significantly different both for small and large values of $\tau$. In particular, in non-geometric models $W$ remains bounded below 1/6.}
% \label{fig:weighted_triangles_models}
% \end{figure*}

It is interesting to observe that the average clustering coefficient can be seen as a specific way of weighing triangles as well:
\begin{equation}\label{eq:nCapprox}
\begin{split}
    &n\overline{C} = \sum_{\substack{i,j,k \in V \\ i<j<k}} \left[\frac{1}{\binom{d_i}{2}} + \frac{1}{\binom{d_j}{2}} + \frac{1}{\binom{d_k}{2}}\right] \mathbbm{1}_{\{i,j,k = \triangle\}}\\
    &\approx n \overline{C}_h : = \sum_{\substack{i,j,k \in V \\ i<j<k}} 2 \left( \frac{1}{h_i^2} + \frac{1}{h_j^2} + \frac{1}{h_k^2} \right) \mathbbm{1}_{\{i,j,k = \triangle\}}. 
\end{split}
\end{equation}
On the other hand,
\begin{equation}\label{eq:Wapprox}
    W \approx W_h := \sum_{\substack{i,j,k \in V \\ i<j<k}} \frac{1}{h_i h_j h_k} \mathbbm{1}_{\{i,j,k = \triangle\}}.
\end{equation}
By~\cite{hofstad2021,blasius2018b,michielan2021}, triangles containing high-weighted vertices are likely to appear in scale-free models regardless of the geometry of the system. Thus, an effective statistic for geometry detection should discount triangles formed by high-degree vertices. From \eqref{eq:nCapprox} and \eqref{eq:Wapprox} we see that such discounting occurs in both $n\overline{C}_h$ and $W_h$. 
However, in $W_h$, the factor $1/(h_i h_j h_k)$ is small as soon as {\it at least one} of the three vertices has high weight. On the contrary,  in $n\overline{C}_h$, the factor $1/h_i^2 + 1/h_j^2 + 1/h_k^2$ is small if and only if {\it all three vertices} have high weights. Therefore, $W$ discounts triangles formed by high-weight vertices more effectively than $n \overline{C}$ does. Indeed, Appendix~\ref{app:2-type_model} shows that the average clustering coefficient may also be constant for non-geometric networks, while weighted triangles are able to detect the non-geometric nature of the underlying network. This makes weighted triangles $W$ a better geometry detector than the average clustering coefficient.

\section{Real-world networks}\label{sec:realnetworks}
We will now investigate the performance of our statistic on real-world data. 
As our statistic was designed to infer hyperbolic geometry, we focus only on networks with a heavy-tailed degree distribution. 
Furthermore, we investigate data sets that contain  multiple snapshots of different sizes of the same network, to compare the theoretical scaling of $W$ with the scaling found in the data. These networks are described in detail in Table \ref{tab:realnetworks}.
\begin{table}[b]
\caption{\label{tab:realnetworks}
Overview of the data sets. \textit{ArXiv collab} collects coauthorships between scientists posting preprints on the Condensed Matter E-Print Archive, from 1999 to 2005 \cite{newman2001}; \textit{CAIDA AS} contains the CAIDA autonomous systems relationship, from January 2004 to November 2007 \cite{caida_as}; \textit{Gnutella p2p} collects all Gnutella peer-to-peer file sharing connections from August 2002 \cite{matei2002}.}
\begin{ruledtabular}
\begin{tabular}{ccc}
\textrm{Name}&
\textrm{\# snapshots}&
\textrm{avg. $n$ per snapshot}\\
\colrule
ArXiv collab \cite{newman2001} & 3 & 29437\\
CAIDA AS \cite{caida_as} & 122 & 22550 \\
Gnutella p2p \cite{matei2002} & 8 & 22811\\
{Bitcoin trans \cite{bitcoin_trans}} & 10 & 3728100\\
\end{tabular}
\end{ruledtabular}
\end{table}

\begin{figure*}[ht]
\centering
\subfloat[ArXiv collab]{\includegraphics[height=10cm]{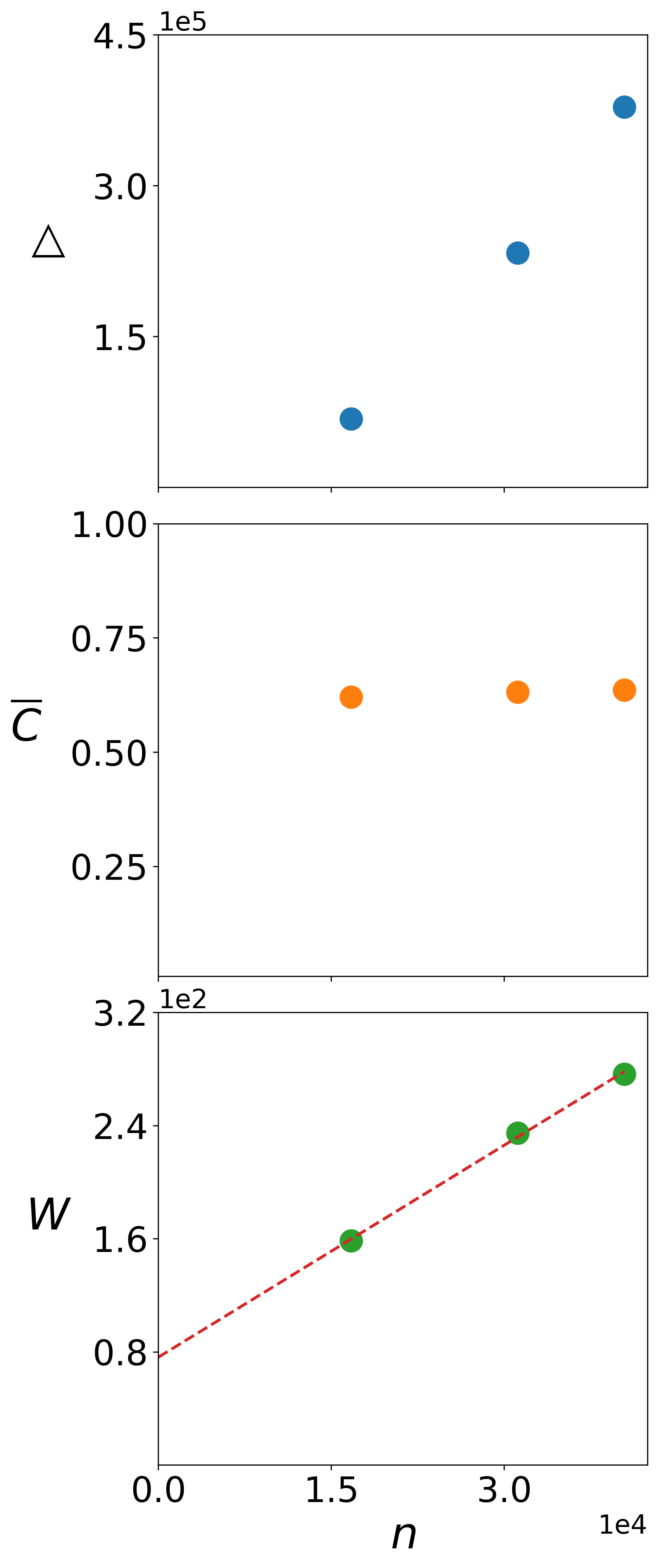}} 
\hspace{0.25cm}
\subfloat[CAIDA AS]{\includegraphics[height=10cm]{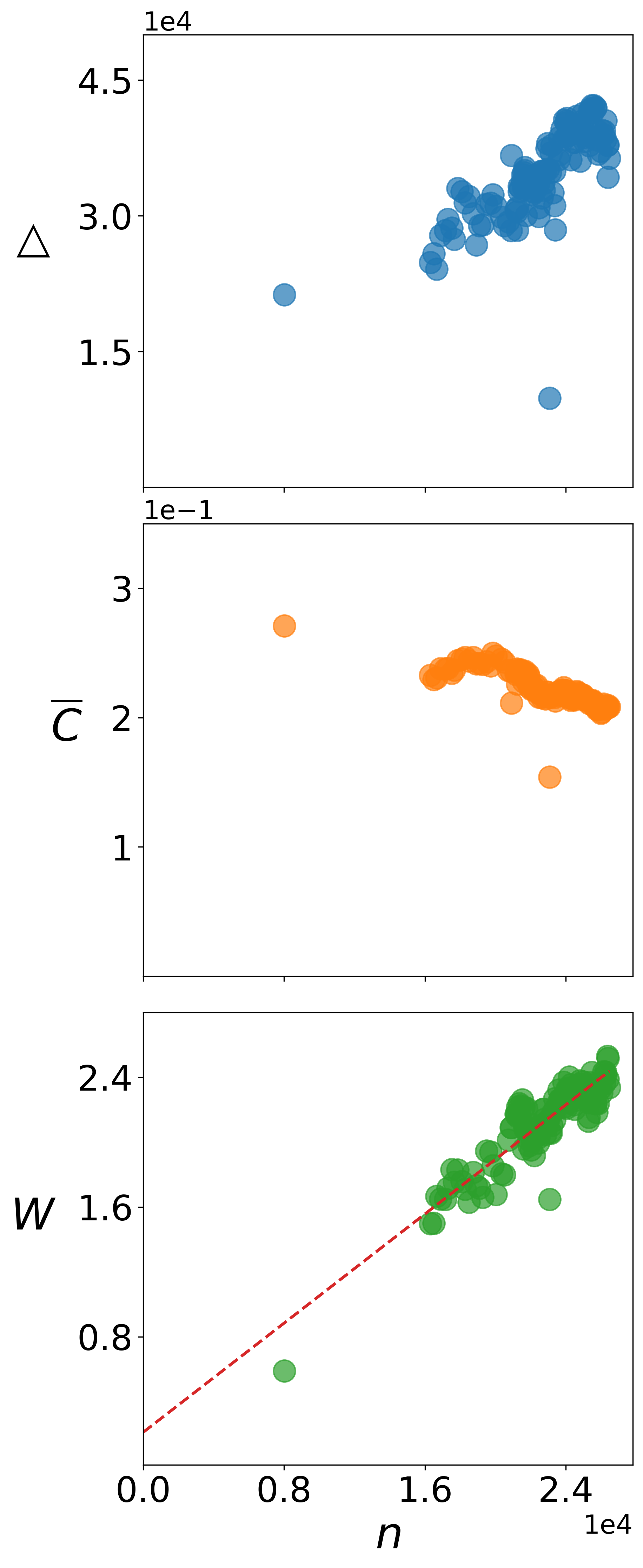}}
\hspace{0.25cm}
\subfloat[Gnutella p2p]{\includegraphics[height=10cm]{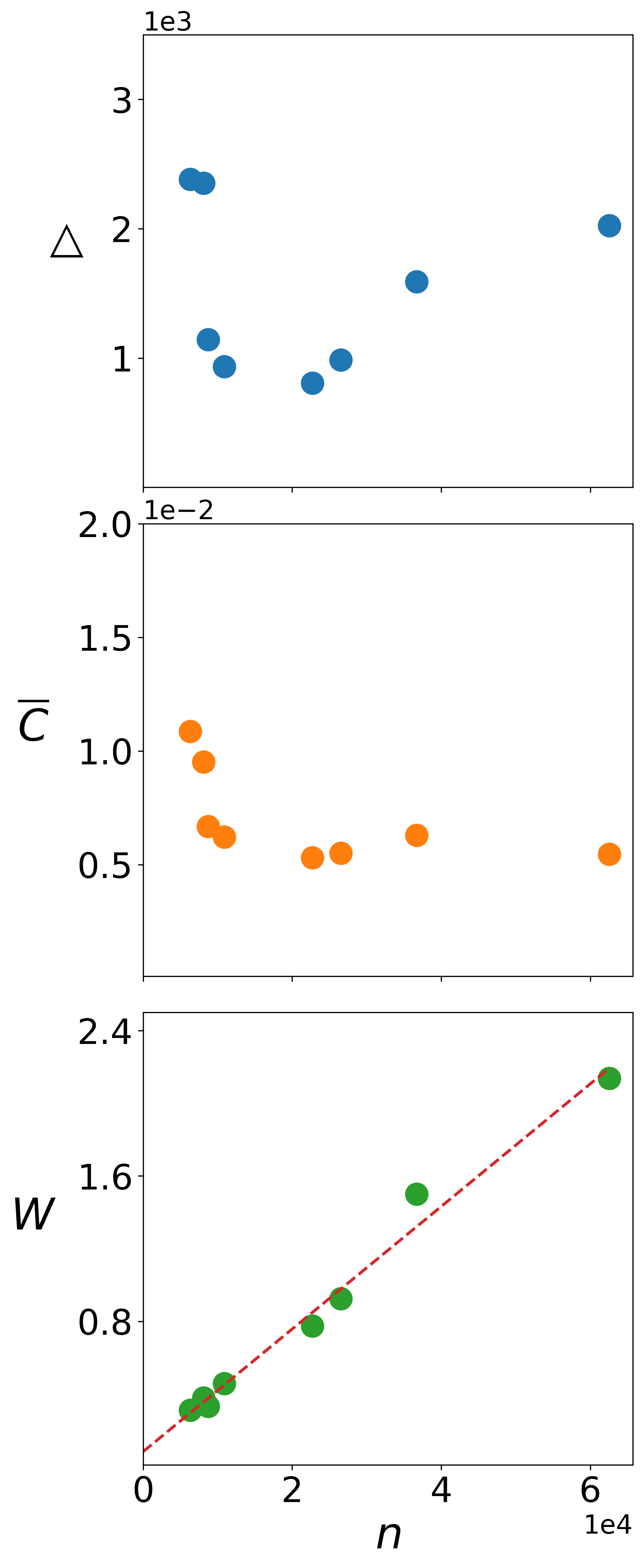}}
\hspace{0.25cm}
\subfloat[Bitcoin trans]{\includegraphics[height=10cm]{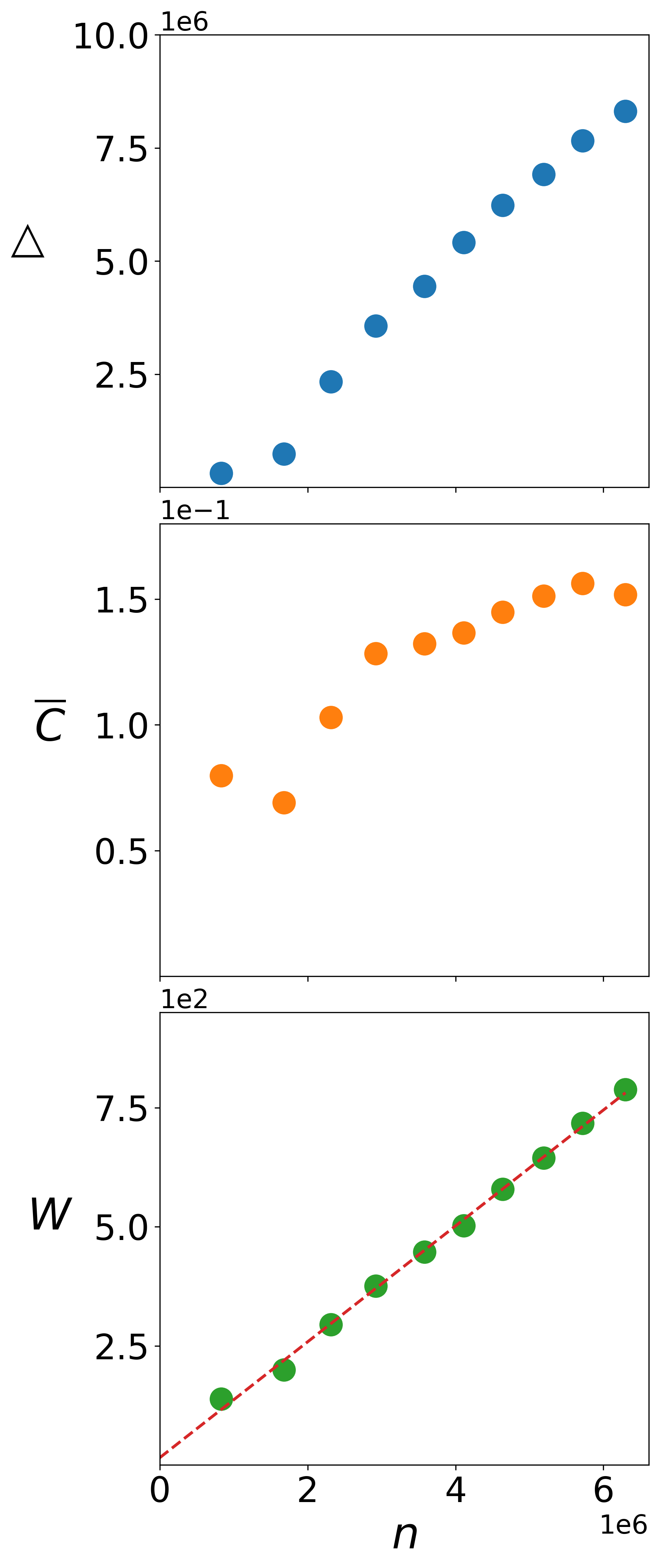}}
\caption{Triangle counts $\triangle$ (blue), average clustering coefficient $\overline{C}$ (orange), and weighted triangles $W$ (green), computed for the data sets: (a) ArXiv collaboration (cond-mat); (b) CAIDA autonomous systems relationship; (c) Gnutella peer-to-peer connections; (d) Bitcoin transactions. The red line is the simple linear regression of $W$.}
\label{fig:real_networks}
\end{figure*}

Figure \ref{fig:real_networks} shows the behavior of triangle counts $\triangle$, the average clustering coefficient $\overline{C}$ and weighted triangles $W$. 

In all four data sets, $W$ behaves roughly linearly in $n$, as our theoretical analysis predicts for geometric networks. The clustering coefficient as well as the pure triangle counts on the other hand, has more erratic behavior that is difficult to interpret. We will now discuss the results for each network in some more detail.

$W$ grows linearly for the ArXiv collaboration networks (Figure \ref{fig:real_networks}a): this suggests that geometry is present in the network. This corresponds to an intuition that ArXiv collaborations have a natural geometric structure: researchers working in the same group, region, or country are more likely to write a paper together. Moreover, a collaboration between two researchers is more likely if they work on similar topics.

The average clustering coefficient $\overline{C}$ is rather small for the CAIDA AS networks (Figure~\ref{fig:real_networks}b), and it seems to be decreasing, as the network size increases. On the other hand, the weighted triangles show a linear growth, revealing a source of geometry, that, in fact, is consistent with the domain knowledge about this network. Indeed, interactions between autonomous systems occur through peering or transit.  An important part of peer selection is the following: internet service providers (ISPs) have a set of \textit{points-of-presence}, that are the locations where they have routers/servers and related equipment/personnel across the world. An ISP can  peer with another ISP only at locations where both have a point-of-presence. Such location-dependent connectivity rules naturally suggest the presence of an implicit geometry in the network formation. 

The Gnutella peer-to-peer networks (Figure \ref{fig:real_networks}c) show an extremely small average clustering coefficient, failing to detect geometry.
On the contrary, the weighted triangles reveal geometry, as $W$ increases linearly with the network size. Indeed, it is reasonable to expect a geometric structure in this network because peer-to-peer file transfers are more likely between users who share similar interests. Then, we could construct an hidden metric space, by embedding users into this space according to their interests.

{Finally, the linear growth of $W$ for the Bitcoin transaction network (Figure \ref{fig:real_networks}d) suggests the presence of geometry in this network as well. This is consistent with {intuition about} the topology of the network: bitcoin transactions have been observed to occur more often between public-key addresses that have entered the network at a similar time \cite{fire2020}.

%We remark that the results presented in this section show that $W$ cannot only be applied to multiple snapshots of one network, but can also be applied to temporal networks.
Note that the Bitcoin transactions network data is a single network with time stamps on the transactions, while the other datasets consist of several snapshots over time. For the Bitcoin network, we therefore create the time-snapshots ourselves by subsampling: we take the induced subgraphs over vertices and edges that have appeared up to a given time. 
In particular, the experiment on the Bitcoin network demonstrates that our method can be applied also to temporal networks which consist of a single issue, if vertices or edges have time stamps. }
%In particular, all real networks of Table \ref{tab:realnetworks} are temporal, and a \textit{time sub-sampling} is used to generate instances of the same network at different sizes.}

Overall, the plots in Figure \ref{fig:real_networks} suggest that weighted triangles can be an effective statistic to detect geometry in real-world networks.

\section{Conclusion and discussion}

In this paper we have analyzed in detail different methods to detect hyperbolic geometry in networks, when only the connections of the network are known. For theoretical purposes, we used the IRG and GIRG as benchmark models, on which we established analytical results. In both random graph models, the probability of connection depends on the popularity features of individuals vertices, expressed through their weights, so that the empirical degree distribution is power law with parameter $\tau$. However, GIRG also models  similarity of individuals, by locating the graph vertices randomly in a $d$-dimensional torus endowed with a metric.

First, we have established that triangle counts and the average clustering coefficient have serious limitations in detecting the presence of geometry when the degree distribution has a particular heavy tail (when $\tau<7/3$). Using numerical simulations we confirmed these limitations, by comparing the number of triangles and the average clustering coefficient in different models.

Next, we have introduced a new statistic, weighted triangles $W$, that effectively discounts triangles formed by high degree vertices. The underlying intuition is that such triangles carry low evidence for geometry because they are common in power law networks with or without an underlying geometric space. Our main  result states that $W$ is a suitable statistic for detecting geometry, because, asymptotically in the network size, $W$ remains bounded in IRGs, while it grows linearly in GIRGs. 

Numerical experiments on the real-world data show a remarkable agreement with this analysis, and confirm the high potential of the $W$ statistic for uncovering a hidden network geometry. 

{
We note that $W$ scales linearly in all four real-life networks presented here. This raises the question whether there exist real-life networks in which $W$ remains bounded. At this point, all non-bipartite real networks we have analyzed seem to possess hidden features which may produce geometric or community effects, such as positions, interests, locations or content similarity. The presence of such features in most real-life networks might explain why we have not found real-world networks with bounded $W$.
}

For practical applications, it would be interesting to know how large $W$ should be to signal geometry. As in IRGs, $\mathbb{E}[W]\leq 1/6$ (Appendix B), Markov's inequality yields that
\begin{equation*}
    \Prob{W \geq n^\alpha} \leq \frac{1}{ n^{\alpha-\varepsilon}} \to 0
\end{equation*}
for any $0<\varepsilon<\alpha<1$. On the other hand, in GIRGs, both $\mathbb{E}[W]$ and $\Var[W]$ are linear in $n$ (see Appendices \ref{app:W_GIRG}-\ref{app:varianceW}). Hence, Chebyshev's inequality gives
\begin{equation*}
    \Prob{W \leq n^{\alpha}} \to 0
\end{equation*}
in the large network limit. This means that $n^\alpha$ would work as a threshold for $W$ to detect geometry, for any $\alpha \in (0,1)$. However, even though {$W$ grows at least linearly} in GIRGs, its leading order term can be small. For example, in Figure \ref{fig:real_networks}c, for $n = 50000$, $W$ is still smaller than 2. Designing statistical tests for finite $n$ is a natural next step, and one may expect that the use of $W$ as test statistic needs to be accompanied by an estimate for the linear slope in GIRGs.

{
{As a potential limitation, we remark that our statistic is designed specifically to} distinguish the geometric GIRG and hyperbolic random graphs from their non-geometric counterparts. {As such, it might not apply to other non-geometric models. For example,} a \textit{household model} \cite{ma2013} does not have an underlying geometry as in GIRG, but it does have {many low-degree triangles within households. We thus believe that weighted triangles will grow linearly in this and other highly clustered non-geometric models. How to distinguish such models from models with  underlying geometry, remains an open question.} }

{
Finally, the growth of $W$ in $n$ can be observed only in temporal networks or snapshots of networks. We illustrated that it {is} possible to give a rule of thumb on how large $W$ should be to signal geometry {in a single-snapshot network, but} this {procedure will} depend on network parameters, such as degree exponents, that {have to be estimated first.} Another {possible approach is} {to artificially generate a series of snapshots of networks of different sizes through} subsampling.   {Subsampling networks while preserving their properties is a difficult problem in itself, and is beyond the scope of this paper. {In the future} it will be interesting to apply weighted triangles to subsamples}  obtained, for example, by methods in randomized algorithms~\cite{orbanz2017}
%We  In \cite{orbanz2017} Orbanz shows how to obtain invariant sub-samples of a network can be obtain via some designed randomized algorithms. Another proposed method to tackle the problem is NetGAN: a pre-trained generative 
or machine-learning \cite{bojchevski2018}. 
%models that create subsampleslearn the distribution of biased random walks on the network \cite{bojchevski2018}. However, the discussion and implementation of these algorithms to obtain sub-samples from single-issue networks goes beyond the scope of this paper.
}

\appendix

\section{Expected degree in GIRG}
\label{app:expgedreeGIRG}
{First, we introduce the asymptotic notation used in the Appendix. Let $f$, $g$ be positive real functions defined on the natural numbers. We say that 
\begin{itemize}
    \item $f(n) = o(g(n))$ when $\lim_{n \to \infty} f(n)/g(n) = 0$;
    \item $f(n) = O(g(n))$  when $\limsup_{n \to \infty} f(n)/g(n) < \infty$;
    \item $f(n) = \Omega(g(n))$ when  $\liminf_{n \to \infty} f(n)/g(n) >0$,
    \item $f(n) = \Theta(g(n))$ when $f(n) = O(g(n))$ and $f(n) = \Omega(g(n))$.
\end{itemize}
We will also consistently use notation $\E_y$ meaning that the averaging occurs with respect to random element $y$. 
}

In this appendix, we calculate the expected degree of a vertex with weight $h_i$ in a GIRG.
First we show that in GIRGs the marginal connection probability, given the weight sequence $(h_i)$, is
\begin{equation*}
    \mathbb{E}_{x_i, x_j}\left[ p(h_i,h_j,x_i,x_j) \right] = \Theta\left(\frac{h_i h_j}{n \mu} \wedge 1 \right),
\end{equation*}
{ where $a \wedge b := \min(a,b)$}.
Indeed, assuming that $x_i$ is fixed and $x_j$ is uniformly distributed, whence $\P(||x_i-x_j||\le r)=\P(||x_j||\le r)$, and denoting $V(r) = \P(||x_j|| \leq r) = (2r)^d$, we obtain
\begin{align*}
    \mathbb{E}_{x_j}[p(h_i,h_j,x_i,x_j)] &=  \int_{[0,1]^d} K_2 \left[\frac{h_i h_j}{n\mu ||x_i-x_j||^d} \wedge 1 \right]^{\gamma} dx_j \\
    &= K_2 \int_0^{1/2} \left[\frac{h_i h_j}{n\mu r^d} \wedge 1 \right]^{\gamma} dV(r) \\
    &= K_2 \int_0^{1/2} \left[\frac{h_i h_j 2^d}{n\mu V(r)} \wedge 1 \right]^{\gamma} dV(r). 
\end{align*}
Then, averaging over $x_i$, and applying integral substitution, we get
\begin{equation*}
\begin{split}
    &\mathbb{E}_{\bx_i}\left[\mathbb{E}_{\bx_j}[p(h_i,h_j,x_i,x_j)]\right] \\
    &= \int_{[0,1]^d} \left(K_2 \int_0^{1} \left[\frac{h_i h_j}{n\mu 2^{-d}V} \wedge 1 \right]^{\gamma} dV \right) dx_i \\
    &= K_2 \int_0^{1} \left[\frac{h_i h_j}{n\mu 2^{-d}V} \wedge 1 \right]^{\gamma} dV.
\end{split}
\end{equation*}
Defining $r_0 := \frac{h_i h_j 2^d}{n \mu}$, consider the next two cases.
\begin{itemize}
    \item If $r_0 \geq 1$, then the latter integrand is always $1$. In such case we conclude $\mathbb{E}_{x_i, x_j}\left[ p(h_i,h_j,x_i,x_j) \right] = 1$.
    \item If $r_0 < 1$, then
    \begin{align*}
        &\mathbb{E}_{x_i, x_j}\left[ p(h_i,h_j,x_i,x_j) \right] \\
        &= K_2 \left(\int_0^{r_0} 1 dV + \int_{r_0}^{1} \left(\frac{r_0}{V} \right)^{\gamma} dV \right)\\
        &=  K_2 \left(r_0 + \left[ \frac{r_0^{\gamma}}{\gamma -1}\left( r_0^{-\gamma + 1} - 1 \right) \right] \right) \\
        &= r_0 K_2 (1 + (\gamma -1)^{-1}) + O(r_0^{\gamma}).
    \end{align*}
    Hence, setting the correction factor to the value $K_2 = (1 - \gamma^{-1}) 2^{-d}$, 
    \begin{multline}\label{eq:marginalgivenw}
        \mathbb{E}_{x_i, x_j}\left[ p(h_i,h_j,x_i,x_j) \right] \\
        = 
        \begin{cases}
        1, & \text{ if $\frac{h_i h_j}{n} \geq \frac{\mu}{2^d}$},\\
        \frac{h_i h_j}{n \mu} + o(\frac{h_i h_j}{n \mu}), & \text{ otherwise}.\\
        \end{cases}
    \end{multline}
\end{itemize} 

Next, we can compute the marginal connection probability of the vertices $i,j$, only given the weight $h_i$. Substituting \eqref{eq:paretolaw} and \eqref{eq:marginalgivenw}, we obtain
\begin{equation*}
    \begin{split}
        \mathbb{E}_{h_j, x_i, x_j}\left[ p(h_i,h_j,x_i,x_j) \right] &= \E_{h_j}[\mathbb{E}_{x_i, x_j}\left[ p(h_i,h_j,x_i,x_j) \right]] \\
        &\approx \int_{\frac{n \mu}{h_i 2^d}}^{\infty}K_1 h^{-\tau} dh \\
        &\hspace{1cm} + \int_{h_0}^{\frac{n \mu}{h_i 2^d}} K_1 \frac{h_i}{n \mu} h^{-\tau+1} dh,
    \end{split}
\end{equation*}
where the approximation follows by dropping the lower order terms. Then:
\begin{equation}
\label{eq:expconnprobGIRG}
\begin{split}
    &\mathbb{E}_{h_j, x_i, x_j}\left[ p(h_i,h_j,x_i,x_j) \right]\\ 
    &\hspace{1cm}= K_1 \frac{h_i h_0^{2-\tau}}{n \mu (\tau-2)} + O\left((h_i/n)^{\tau - 1}\right)\\
    &\hspace{1cm}= \frac{h_i}{n} (1 + o(1)),
\end{split}
\end{equation}
where the latter equality follows from the fact that $K_1 = (\tau-1)h_0^{\tau-1}$, $\mu = \frac{\tau-1}{\tau-2}h_0$.
In other words, Equation \eqref{eq:expconnprobGIRG} tells that a vertex $i$ with weight $h_i$ connects to any other vertex with probability $h_i/n$, in the large $n$ limit.

Finally, the expected degree of a vertex $i$ with weight $h_i$ is
\begin{equation*}
\begin{split}
    \E[\text{deg}(i)] &= \E\left[\sum_{j \neq i} \mathbbm{1}_{\{(i \leftrightarrow j | h_i)\}} \right] \\
    &= (n-1)\E_{h_j,x_i,x_j}[p(h_i,h_j,x_i,x_j)],
\end{split}
\end{equation*}
and from Equation \ref{eq:expconnprobGIRG} we conclude that
\begin{equation*}
    \E[\text{deg}(i)] \approx h_i.
\end{equation*}

\section{Upper bound for $\E[W]$ in IRGs}\label{app:W_IRG}
The expected value of $W$ in a IRG is given by
\begin{equation}
    \E[W] = \sum_{i<j<k} \E\left[\frac{1}{d_i d_j d_k} \mathbbm{1}_{\{(i,j,k)=\triangle\}}\right]\label{eq:expectedWIRG}.
\end{equation}
To obtain the order of magnitude of \eqref{eq:expectedWIRG}, we compute the upper bound of a similar quantity, where the degrees are replaced by the corresponding independent weights. In other words, $\E[W]$ is roughly approximated by
\begin{multline*}
   \E[W_h]=\binom{n}{3} \int_{[h_0,\infty]^3} (h_1 h_2 h_3)^{-1} \prod_{1 \leq \alpha < \beta \leq 3} \left(\frac{h_{\alpha} h_{\beta}}{\mu n} \wedge 1 \right) \\ \times \rho(h_1) \rho(h_2) \rho(h_3) \; d h_1 \; d h_2\; d h_3,
\end{multline*}
where the binomial term arises from counting the possible combination of three vertices in $V$.

Since $\left(\frac{h_{\alpha} h_{\beta}}{\mu n} \wedge 1 \right) \leq \frac{h_{\alpha} h_{\beta}}{\mu n}$ for all $\alpha, \beta$, the latter integral is upper bounded by
\begin{equation*}
    \left(\frac{K_1}{\mu n}\right)^3 \int_{[h_0,\infty]^3} (h_1 h_2 h_3)^{1 - \tau} dh_1 dh_2 dh_3, \label{eq:upboundintWIRG} 
\end{equation*}
where $K_1$ is given by the Pareto power-law distribution for the weights. Then,
\begin{align*}
    \label{eq:eWIRGresult}
    \E[W_h]\ &\leq \binom{n}{3} \left(\frac{K_1}{\mu n}\right)^3 \left(\int_{h_0}^{\infty} h^{1-\tau} dh \right)^3 \\
    & = \binom{n}{3} \left(\frac{K_1}{\mu n}\right)^3 \left(\frac{h_0^{-\tau+2}}{\tau-2} \right)^3 \\
    &=  \binom{n}{3} \frac{1}{n^3} \leq \frac{1}{6},
\end{align*}
where in the second equality we used the fact that $K_1/\mu = (\tau-2)h_0^{\tau-2}$.
Since the degrees $d_i$ in the IRG are concentrated around their means $h_i$, we conclude that $\E[W]=O(1)$.

\section{Lower bound for $\E[W]$ in GIRGs}\label{app:W_GIRG}

The expected value of $W$ in GIRGs is 
\begin{equation}
    \E[W] = \sum_{i,j,k \in V} \E\left[\frac{1}{d_i d_j d_k} \mathbbm{1}_{\{(i,j,k)=\triangle\}}\right]\label{eq:expectedWGIRG}.
\end{equation}
In a similar way as in Appendix \ref{app:W_IRG}, to obtain the order of magnitude of \eqref{eq:expectedWGIRG} we lower bound a similar quantity, where the degrees are replaced by the corresponding independent weights. That is, $\E[W]$ is roughly approximated by

\begin{widetext}
\begin{equation}\label{eq:expectedWGIRGint}
    \E[W_h] = \binom{n}{3}\int_{[h_0,\infty]^3} K_1^3(h_1 h_2 h_3)^{-\tau-1} \int_{[0,1]^{3d}} \prod_{1 \leq \alpha < \beta \leq 3} K_2\left(\frac{h_{\alpha} h_{\beta}}{\mu n ||x_{\alpha} - x_{\beta}||^d} \wedge 1 \right)^{\gamma} \; dx_1 \; dx_2 \; dx_3 \; dh_1 \; dh_2\; dh_3.
\end{equation}
\end{widetext}

Observe that 
%\begin{equation*}
%    \begin{split}
%        ||x_2 - x_3|| &\leq ||x_2 - x_1|| + ||x_1 - %x_3|| \\
%        &\leq 2 \max\{||x_1 - x_3||, ||x_1 - x_2||\},   
%    \end{split}
%\end{equation*} 
$||x_2 - x_3|| \leq 2 \max\{||x_1 - x_3||, ||x_1 - x_2||\}$
by the triangle inequality. 

Then
\begin{equation*}
    \begin{split}
        \frac{1}{||x_2 - x_3||^d} &\geq \frac{1}{2^d \max\{||x_1 - x_3||^d, ||x_1 - x_2||^d\}} \\&= \left( \frac{1}{2^d ||x_1 - x_2||^d} \wedge \frac{1}{2^d ||x_1 - x_3||^d} \right).
    \end{split}
\end{equation*}
 Substituting $(x_1,x_2,x_3) \to (z_1,z_2,z_3):=(x_1, x_2 - x_1, x_3 - x_1)$, the integral in \eqref{eq:expectedWGIRGint} is bounded from below by
\begin{multline}
    \int_{[h_0,\infty]^3} K_1^3 (h_1 h_2 h_3)^{-\tau-1}  \\ \times K_2^3 \int_{[-1/2,1/2]^{2d}} \left(\frac{h_{2} h_{3}}{\mu n 2^d ||z_2||^d} \wedge \frac{h_{2} h_{3}}{\mu n 2^d ||z_3||^d} \wedge 1 \right)^{\gamma} \\ \times \left(\frac{h_{1} h_{2}}{\mu n ||z_2||^d} \wedge 1 \right)^{\gamma} \left(\frac{h_{1} h_{3}}{\mu n ||z_3||^d} \wedge 1 \right)^{\gamma} \\ 
    \times dz_2 \; dz_3 \; dh_1 \; dh_2\; dh_3. \label{eq:threeintegrands}
\end{multline}
Now, consider the cube 
\begin{equation*}
    \mathcal{C} := \left[-\left(\frac{h_0^2}{\mu n 2^d}\right)^{1/d}, \left(\frac{h_0^2}{\mu n 2^d}\right)^{1/d}\right]^d.
\end{equation*}
If $z_2 \in \mathcal{C}$ and $z_3 \in \mathcal{C}$, then all the minima in the integrand of \eqref{eq:threeintegrands} are equal to 1. Therefore, after restricting the spatial integrals to the subset $\mathcal{C}^2$, \eqref{eq:threeintegrands} is lower bounded by
\begin{equation*}
    \int_{[h_0,\infty]^3} (K_1 K_2)^3 (h_1 h_2 h_3)^{-\tau-1} \; dh_1 \; dh_2 \; dh_3 \int_{\mathcal{C}^2} dz_2 \; dz_3.
\end{equation*}
Then, we obtain
\begin{equation*}
    \begin{split}
        \E[W_h] &\geq \binom{n}{3} \left(\int_{h_0}^{\infty} K_1 K_2 h ^{-\tau-1} \; dh \right)^3 \left(\int_{\mathcal{C}^2} dz \right)^2 \\
    &= \binom{n}{3} (K_1 K_2)^3 \frac{h_0^{3\tau}}{\tau^3}\left(\frac{h_0^2}{\mu n}\right)^2\\
    &= K_3 n (1 + o(1)),
    \end{split}
\end{equation*}
where $K_3$ is a non-null constant depending on the GIRG parameters $\tau, h_0, \gamma, d$. 
Since the degrees $d_i$ in the GIRG are
concentrated around their means $h_i$, we conclude that
$\E[W] = \Omega(n)$.

\section{Upper bound for $\Var[W]$ in GIRGs}\label{app:varianceW}

Observe that
\begin{align}
\begin{split}
    &\Var[W_h] = \Var\left[ \sum_{i<j<k} \frac{\mathbbm{1}_{\{i,j,k = \triangle\}}}{h_i h_j h_k} \right]\\
    &= \sum_{\substack{i_1<j_1<k_1 \\ i_2<j_2<k_2}} \Cov \left( \frac{\mathbbm{1}_{\{i_1,j_1,k_1 = \triangle\}}}{h_{i_1}h_{j_1}h_{k_1}} ,  \frac{\mathbbm{1}_{\{i_2,j_2,k_2 = \triangle\}}}{h_{i_2}h_{j_2}h_{k_2}} \right) \\
    &\leq\sum_{\substack{i_1<j_1<k_1 \\ i_2<j_2<k_2}} \Exp{\frac{\mathbbm{1}_{\{i_1,j_1,k_1 = \triangle\}} \mathbbm{1}_{\{i_2,j_2,k_2 = \triangle\}}}{h_{i_1}h_{j_1}h_{k_1}h_{i_2}h_{j_2}h_{k_2}}}\\
    &= \sum_{\substack{i_1<j_1<k_1 \\ i_2<j_2<k_2}} \mathbb{E}_{\overline{h},\overline{x}} \left[\frac{\Prob{i_1,j_1,k_1 = \triangle}\Prob{i_2,j_2,k_2 = \triangle}}{h_{i_1}h_{j_1}h_{k_1}h_{i_2}h_{j_2}h_{k_2}}\right], 
\end{split}\label{eq:varboundWGIRG}
\end{align}
where indices $\overline{h},\overline{x}$ mean that the expectation is performed over all the random variables $(h_i)_{i \in [n]}$ and $(x_i)_{i \in [n]}$.
There are now 4 different cases for the choices of $i_1,j_1,k_1,i_2,j_2,k_2$:
\begin{enumerate}[(i)]
    \item $(i_1,j_1,k_1)$ and $(i_2,j_2,k_2)$ do not intersect.\\
    In this case the covariance is 0, because the two random variables $\frac{\mathbbm{1}_{\{i_1,j_1,k_1 = \triangle\}}}{h_{i_1}h_{j_1}h_{k_1}}$ and $\frac{\mathbbm{1}_{\{i_2,j_2,k_2 = \triangle\}}}{h_{i_2}h_{j_2}h_{k_2}}$ are independent.
    
    \item $(i_1,j_1,k_1)$ and $(i_2,j_2,k_2)$ intersect at 1 vertex (wlog $i_1 = i_2$).
    There are $O(n^5)$ ways to choose the vertices. We can bound the probabilities
    \begin{equation*}
    \begin{split}
        \Prob{i_1,j_1,k_1 = \triangle} &\leq p(h_{i_1},h_{j_1},x_{i_1},x_{j_1})p(h_{j_1},h_{k_1},x_{j_1}, x_{k_1})\\
        \Prob{i_2,j_2,k_2 = \triangle} &\leq p(h_{i_2},h_{j_2},x_{i_2},x_{j_2})p(h_{j_2},h_{k_2},x_{j_2}, x_{k_2})
    \end{split}
    \end{equation*}
    Moreover, for any sequence of nodes $(\alpha_1,...,\alpha_m)$,
    \begin{multline*}
        \mathbb{E}_{\overline{x}}\left[\prod_{\ell=1}^{m-1} p(h_{\alpha_\ell},h_{\alpha_{\ell+1}},x_{\alpha_{\ell} },x_{\alpha_{\ell+1}}) \right] \\ = \prod_{\ell=1}^{m-1} \mathbb{E}_{\overline{x}}\left[p(h_{\alpha_\ell},h_{\alpha_{\ell+1}},x_{\alpha_{\ell}},x_{\alpha_{\ell+1}}) \right]. 
    \end{multline*}
    This is a consequence of the fact that the positions are i.i.d. uniformly distributed in the torus, and that the connection probability is a function of the distance.
    Then, the terms in the sum of the r.h.s. in \eqref{eq:varboundWGIRG} are upper bounded by
    \begin{widetext}
    \begin{multline*}
        \mathbb{E}_{\overline{h}}\left[ \frac{\E_{\overline{x}}[p(h_{k_1},h_{j_1},x_{k_1},x_{j_1})]\E_{\overline{x}}[p(h_{j_1},h_{i_1},x_{j_1},x_{i_1})]\E_{\overline{x}}[p(h_{i_1},h_{j_2},x_{i_1},x_{j_2})]\E_{\overline{x}}[p(h_{j_2},h_{k_2},x_{j_2},x_{k_2})]}{h_{i_1}^2h_{j_1}h_{k_1}h_{j_2}h_{k_2}}\right] \\
        = \int_{[h_0,\infty)^5} K_1^5  h_{i_1}^{-\tau-2}  (h_{j_1} h_{k_1} h_{j_1} h_{k_2})^{-\tau -1} \frac{h_{k_1} h_{j_1}^2 h_{i_1}^2 h_{j_2}^2 h_{k_2}}{(n \mu)^4} \; dh_{i_1} \; dh_{j_1} \; dh_{k_1} \; dh_{j_2} \; dh_{k_2} = \frac{1}{n^4 \mu^2}.
    \end{multline*}
    \end{widetext}
    where the first equality follows from \eqref{eq:marginalgivenw}.
    Then, the contribution to the variance from triangles overlapping in one vertex is $O(n^5/n^4) = O(n)$.
    
    \item $(i_1,j_1,k_1)$ and $(i_2,j_2,k_2)$ do intersect in 2 vertices (wlog $i_1 = i_2, j_1 = j_2$).\\
    In this case the number of possible choices for the vertices is $O(n^4)$. We may bound the term in the r.h.s. of \eqref{eq:varboundWGIRG} with
    \begin{widetext}
    \begin{multline*}
        \mathbb{E}_{h}\left[ \frac{\E_{\overline{x}}[p(h_{k_1},h_{j_1},x_{k_1j_1})]\E_{\overline{x}}[p(h_{j_1},h_{i_1},x_{j_1i_1})]\E_{\overline{x}}[p(h_{i_1},h_{k_2},x_{i_1k_2})]}{h_{i_1}^2h_{j_1}^2h_{k_1}h_{k_2}}\right] \\
        = \int_{[h_0,\infty)^4} K_1^4  (h_{i_1}h_{j_1})^{-\tau-2} (h_{k_1}h_{k_2})^{-\tau -1} \frac{h_{k_1} h_{j_1}^2 h_{i_1}^2 h_{k_2}}{(n \mu)^3} \; dh_{i_1} \; dh_{j_1} \; dh_{k_1} \; dh_{k_2} = \frac{1}{n^3 \mu^3},
    \end{multline*}
    \end{widetext}
    where the first equality follows from \eqref{eq:marginalgivenw}. Then, the contribution to the variance from triangles overlapping in two vertices is $O(n^4/n^3) = O(n)$.
    
    \item $(i_1,j_1,k_1)$ and $(i_2,j_2,k_2)$ intersect in 3 vertices ($i_1 = i_2, j_1 = j_2, k_1 = k_2$).\\
    There are $\binom{n}{3}$ ways to choose the vertices. We upper bound the term in the r.h.s. of \eqref{eq:varboundWGIRG} with
    \begin{widetext}
    \begin{equation*}
        \mathbb{E}_{h}\left[ \frac{\E_{\overline{x}}[p(h_{k_1},h_{j_1},x_{k_1j_1})]\E_{\overline{x}}[p(h_{j_1},h_{i_1},x_{j_1i_1})]}{h_{i_1}^2h_{j_1}^2h_{k_1}^2} \right] = \int_{[h_0,\infty)^3} K_1^3  (h_{i_1}h_{j_1}h_{k_1})^{-\tau-2} \frac{h_{i_1} h_{j_1}^2 h_{k_1}}{(n \mu)^2} \; dh_{i_1} \; dh_{j_1} \; dh_{k_1} = \frac{1}{n^2} \left(\frac{\tau-2}{\tau h_0^2}\right)^2,
    \end{equation*}
    \end{widetext}
    where, again, the first equality follows from \eqref{eq:marginalgivenw}. Then, the contribution from triangles completely overlapping is $O(n)$.
\end{enumerate}
Hence, summing up all possible different cases, we see that the variance of $W_h$ in GIRGs is $O(n)$. Since the degrees $d_i$ in the GIRG
are concentrated around their means $h_i$, we conclude that $\Var[W] = O(n)$.

\section{Concentration of $W$}\label{app:concentrationW}
Here we show that $W$ concentrates around its expected value both in IRGs and in GIRGs.

In Appendix \ref{app:W_IRG} we proved that $\E[W] = O(1)$, in IRGs. Then,
\begin{equation}\label{eq:markovWIRG}
    W = O(1), \quad \text{with high probability}
\end{equation}
in IRG, as consequence of the Markov's inequality.

On the other hand, we proved that $\E[W] = \Omega(n)$ and $\Var[W] = O(n)$ in GIRGs (Appendices \ref{app:W_GIRG}-\ref{app:varianceW}). By Chebyshev's inequality,
\begin{equation}\label{eq:chebyshevWGIRG}
    \Prob{\left|\frac{W}{\E[W]} - 1 \right| > \delta} \leq \frac{\Var[W]}{\delta^2 \E[W]^2} = o(1),
\end{equation}
for all $\delta > 0$. In particular, from Equation \eqref{eq:chebyshevWGIRG} follows that, in GIRGs, for all $\varepsilon > 0$, there exist $\overline{\delta} > 0$, $n_0 > 0$ such that $\Prob{W < \overline{\delta} n }<\varepsilon$, for all $n>n_0$. That is,
\begin{equation}
    W = \Omega(n), \quad \text{with high probability}.
\end{equation}

\section{2-type inhomogeneous model}\label{app:2-type_model}
Consider an inhomogeneous random graph where 
\begin{equation*}
    h_i=\begin{cases}
    2, & \text{with probability }1-1/(\sqrt{\mu n}),\\
    \sqrt{\mu n}, & \text{with probability } 1/(\sqrt{\mu n}).
    \end{cases}
\end{equation*}

We can approximate the weighted triangles $W$ by
\begin{equation*}
    W_h=\sum_{i,j,k}\triangle_{i,j,k}\big(h_i h_j h_k\big)^{-1}.
\end{equation*}

Similarly, we can approximate the average clustering coefficient $\overline{C}$ by
\begin{equation*}
    n \overline{C}_h=\sum_{i,j,k}\triangle_{i,j,k}\big(h_i^{-2}+h_j^{-2}+h_k^{-2}\big).
\end{equation*}

In this graph, a triangle falls into one of the four categories:
\begin{enumerate}[(I)]
    \item The triangle has vertices with $h_i=h_j=h_k=2$. \\
    Each one of these triangles contributes 
    \begin{equation*}
        h_i^{-2}+h_j^{-2}+h_k^{-2} = 3/4
    \end{equation*}
    to $\overline{C}$, and 
    \begin{equation*}
        (h_i h_j h_k)^{-1} = 1/8
    \end{equation*}
    to $W_h$.\\
    There are $O(n^3)$ sets containing 3 vertices of weight 2, and they form a triangle with probability 
    \begin{equation*}
    \begin{split}
        &p(h_i,h_j)p(h_i,h_k)p(h_j,h_k) \\
        &\hspace{1cm}= \left(\frac{4}{n \mu} \wedge 1 \right) \left(\frac{4}{n \mu} \wedge 1 \right) \left(\frac{4}{n \mu} \wedge 1 \right) \\
        &\hspace{1cm}= O(1/n^3). 
    \end{split}
    \end{equation*}
    Thus, this triangle type contributes $O(1)$ to $n \overline{C}_h$, and $O(1)$ to $W_h$. 
    \item The triangle has $h_i=h_j=2$, $h_k=\sqrt{\mu n}$.\\
    Each one of these triangles contributes 
    \begin{equation*}
        h_i^{-2}+h_j^{-2}+h_k^{-2} = \frac{1}{2} + \frac{1}{n \mu} \approx \frac{1}{2}
    \end{equation*}
    to $\overline{C}$, and 
    \begin{equation*}
        (h_i h_j h_k)^{-1} = \frac{1}{4 \sqrt{n \mu}}
    \end{equation*}
    to $W_h$.\\
    There are $\Theta(n^{5/2})$ sets containing 2 vertices of weight 2 and one with weight $\sqrt{\mu n}$, and they form a triangle with probability 
    \begin{equation*}
    \begin{split}
        &p(h_i,h_j)p(h_i,h_k)p(h_j,h_k) \\
        &\hspace{1cm}= \left(\frac{4}{n \mu} \wedge 1 \right) \left(\frac{2}{\sqrt{n \mu}} \wedge 1 \right) \left(\frac{2}{\sqrt{n \mu}} \wedge 1 \right) \\
        &\hspace{1cm}= \Theta(n^{-2}). 
    \end{split}
    \end{equation*}
    Thus, this triangle type contributes $\Theta(\sqrt{n})$ to $n \overline{C}_h$, and $O(1)$ to $W_h$. 
    \item The triangle has $h_i=2$, $h_j=h_k=\sqrt{\mu n}$.\\
    Each one of these triangles contributes 
    \begin{equation*}
        h_i^{-2}+h_j^{-2}+h_k^{-2} = \frac{1}{4} + \frac{2}{n \mu} \approx \frac{1}{4}
    \end{equation*}
    to $\overline{C}$, and 
    \begin{equation*}
        (h_i h_j h_k)^{-1} = \frac{1}{2 n \mu}
    \end{equation*}
    to $W_h$.\\
    There are $\Theta(n^{2})$ sets containing of one vertex of weight 2 and two with weight $\sqrt{\mu n}$, and they form a triangle with probability 
    \begin{equation*}
    \begin{split}
        &p(h_i,h_j)p(h_i,h_k)p(h_j,h_k) \\
        &\hspace{1cm}= \left(\frac{2}{\sqrt{n \mu}} \wedge 1 \right) \left(\frac{2}{\sqrt{n \mu}} \wedge 1 \right)\\
        &\hspace{1cm}= \Theta(n^{-1}). 
    \end{split}
    \end{equation*}
    Thus, this triangle type contributes $\Theta(n)$ to $n \overline{C}_h$, and $O(1)$ to $W_h$. 
    \item The triangle has $h_i=h_j=h_k=\sqrt{\mu n}$. \\
    Each one of these triangles contributes 
    \begin{equation*}
        h_i^{-2}+h_j^{-2}+h_k^{-2} = \frac{3}{n \mu}
    \end{equation*}
    to $\overline{C}$, and 
    \begin{equation*}
        (h_i h_j h_k)^{-1} = \frac{1}{(n \mu)^3/2}
    \end{equation*}
    to $W_h$.\\
    There are $\Theta(n^{3/2})$ sets containing three vertices with weight $\sqrt{\mu n}$, and they form a triangle with probability 
    \begin{equation*}
    \begin{split}
        p(h_i,h_j)p(h_i,h_k)p(h_j,h_k) = 1. 
    \end{split}
    \end{equation*}
    Thus, this triangle type contributes $\Theta(\sqrt{n})$ to $n \overline{C}_h$, and $O(1)$ to $W_h$. 
\end{enumerate}

Thus the total contributions to $n \overline{C}$ are:
\begin{itemize}
    \item $O(1)$, from type (I) triangles;
    \item $\Theta(\sqrt{n})$, from type (II) triangles;
    \item $\Theta(n)$, from type (III) triangles;
    \item $\Theta(\sqrt{n})$, from type (IV) triangles.
\end{itemize}
Hence, $n \overline{C}_h = \Theta(n)$.

On the other hand, the total contributions to $W$ are:
\begin{itemize}
    \item $O(1)$, from type (I) triangles;
    \item $\Theta(1)$, from type (II) triangles;
    \item $\Theta(1)$, from type (III) triangles;
    \item $\Theta(1)$, from type (IV) triangles.
\end{itemize}
Hence, $W_h = \Theta(1)$.

\begin{comment}
\section{General $f$ conditions}
We define our general function as
\begin{equation}
    \sum_{i,j,k}\mathbbm{1}_{\triangle_{i,j,k}}f(d_i,d_j,d_k),
\end{equation}
where $f(x,y,z)$ is a symmetric function. Now under the conditions
\begin{equation}
    \int_{1}^{\infty}\int_{1}^{\infty}\int_{1}^{\infty}(h_1h_2h_3)^{2-\tau}f(h_1,h_2,h_3) dh_1dh_2dh_3<\infty,
\end{equation}
and 
\begin{equation}
    \int_{1}^{\infty}\int_{1}^{\infty}\int_{1}^{\infty}(h_1h_2h_3)^{-\tau}f(h_1,h_2,h_3) dh_1dh_2dh_3>0,
\end{equation}
our conclusions still hold. This encompasses a wide class of weighted clustering-based statistics that enable to detect hyperbolic geometry. In particular, functions such as $f(x,y,z)=(xyz)^{-\delta}$ for $\delta\geq 1$, $f(x,y,z)=\mathbbm{1}_{x<M}\mathbbm{1}_{y<M}\mathbbm{1}_{z<M}$ for any $M>1$ or $f(x,y,z)=\exp(-\alpha x)\exp(-\alpha y)\exp(-\alpha z)$ for any $\alpha>0$ enable to detect geometry. 
\end{comment}

%\nocite{*}

\bibliography{biblio}% Produces the bibliography via BibTeX.

\end{document}